\documentclass[fleqn,twoside]{article}

\usepackage{espcrc2}
\usepackage{graphicx}

\def\simgt{\,\rlap{\lower 3.5 pt\hbox{$\mathchar \sim$}}\raise 1pt \hbox {$>$}\,}
\def\simlt{\,\rlap{\lower 3.5 pt\hbox{$\mathchar \sim$}}\raise 1pt \hbox {$<$}\,}

\title{
Chiral extrapolations with small sea quark mass data
in two-flavor lattice QCD
\thanks{Talk presented by Y.~Namekawa}
}

\author{CP-PACS Collaboration:
  Y.~Namekawa\rlap,\address{Grad. School of Pure and Appl. Sciences,
       University of Tsukuba, Tsukuba, Ibaraki 305-8571, Japan}
  \thanks{Present address:
      Department of Physics,
      Nagoya University,
      Nagoya 464-8602, Japan
 }
  S.~Aoki\rlap,$^{a}$
  M.~Fukugita\rlap,\address{Institute for Cosmic Ray Research,
      University of Tokyo, Kashiwa 277-8582, Japan}
  K-I.~Ishikawa\rlap,$^{a,}$\address{Center for Computational Sciences,
      University of Tsukuba, Tsukuba, Ibaraki 305-8577, Japan}
  \thanks{Present address:
      Department of Physics,
      Hiroshima University,
      Higashi-Hiroshima, Hiroshima 739-8526, Japan
 }
  N.~Ishizuka\rlap,$^{a,c}$
  Y.~Iwasaki\rlap,$^{a}$
  K.~Kanaya\rlap,$^{a}$
  T.~Kaneko\rlap,\address{High Energy Accelerator Research Organization
      (KEK), Tsukuba, Ibaraki 305-0801, Japan}
  Y.~Kuramashi\rlap,$^{d}$%
  \thanks{Present address:
      Center for Computational Physics,
      University of Tsukuba, Tsukuba, Ibaraki 305-8577, Japan
 }
  V.I.~Lesk\rlap,$^{c}$%
  \thanks{Present address:
      Department of Biological Sciences,
      Imperial College, London SW7 2AZ, U.K.
 }
  M.~Okawa\rlap,\address{Department of Physics,
      Hiroshima University,
      Higashi-Hiroshima, Hiroshima 739-8526, Japan}
  A.~Ukawa$^{a,c}$
  T.~Umeda\rlap,$^{c}$%
  \thanks{Present address:
      Yukawa Institute for Theoretical Physics,
      Kyoto University, Kyoto 606-8502, Japan
 }
  and
  T.~Yoshi\'e$^{a,c}$
 }

\begin{document}

\begin{abstract}
We present results on the light hadron spectrum
and quark mass in two-flavor QCD calculated
with small sea quark masses down to $m_{PS}/m_{V}=0.35$.
The configurations are generated
using the RG improved gauge
and tadpole-improved clover quark action
at $\beta=1.8$, where $a^{-1} \simeq 1$~GeV.
Chiral extrapolations are made using not only polynomials
and ChPT in the continuum but also formulae of Wilson chiral
perturbation theory (WChPT) including $O(a^2)$ chiral breaking terms.
We examine the viability of WChPT and its influence on quark masses.
\end{abstract}

\maketitle
\thispagestyle{empty}


\section{Introduction}

Systematic errors associated with chiral extrapolations
present a serious obstacle to precise QCD calculations~\cite{Ishikawa}.
A rapid increase of the computational cost toward light quarks
limits quark masses to the values larger than their physical ones,
especially for dynamical Wilson-type quarks. 
This limitation causes sizable ambiguities 
in the extrapolation to the physical point.

In this report, we analyze issues related to small quark masses and 
finite lattice spacings in the chiral extrapolation using   
several chiral extrapolation functions.  
Our study is based on hadron mass data for small sea quark masses  
in the range $m_{PS}/m_{V}=0.60-0.35$~\cite{Namekawa},
combined with our previous data at heavier quark masses at
$m_{PS}/m_{V}=0.80-0.55$~\cite{CP-PACS.Nf2}. 
These hadron mass data are generated with 
an RG-improved gauge 
action and a meanfield-improved clover quark action
at $\beta=1.8$ ($a \simeq 0.2$~fm).
We use $12^{3} \times 24$ and $16^{3} \times 24$ lattices 
with the spatial size of 2.4 and 3.2~fm. 
Our simulation parameters
are summarized in Table~\ref{table:simulation_parameters}. 
For further details, see Ref.~\cite{Namekawa}.


\begin{table}[htbp]
\begin{center}
\begin{tabular}{llll} 
\hline
$\kappa_{sea}$&$(m_{PS}/m_{V})_{sea}$&\multicolumn{2}{c}{$N_{traj}$}\\
&&$12^3\times 24$&$16^3\times 24$\\
\hline
0.1409 & 0.807(1) & 6250\\
0.1430 & 0.753(1) & 5000\\
0.1445 & 0.694(2) & 7000\\
0.1464 & 0.547(4) & 5250\\
\hline
0.14585 & 0.609(2) & 4000\\
        & 0.604(3) &      & 2000\\
0.14660 & 0.509(5) & 4000\\
        & 0.509(4) &      & 2000\\
0.14705 & 0.413(8) & 4000\\
0.14720 & 0.349(19) & 1400\\
\hline
\end{tabular}
\end{center}
\caption{Simulation parameters at $\beta=1.8$. The top four runs are 
from Ref.~\cite{CP-PACS.Nf2}.}
\label{table:simulation_parameters}
\end{table}


\section{Polynomial extrapolation}

\begin{figure}[tb]
\vspace{-6mm}
\begin{center}
\scalebox{0.60}{\includegraphics{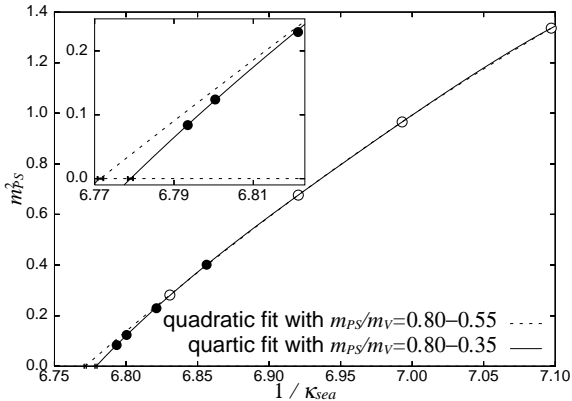}}
\end{center}
\vspace{-13mm}
\caption{
      Chiral extrapolation of pseudoscalar meson mass.
      Open symbols show the results obtained
      in our previous calculation \protect\cite{CP-PACS.Nf2}.
}
\label{figure:kappa_inv-m_PS2}
\vspace{-6mm}
\end{figure}

The conventional polynomial extrapolation has a problem that
polynomials are not consistent with the logarithmic singularity
expected in the chiral limit.
It is imperative to estimate the systematic errors
due to higher order contributions
when the data are extrapolated using a low-order polynomial.
Figure~\ref{figure:kappa_inv-m_PS2} shows that
our new data at small sea quark masses
deviate systematically from the quadratic fit of old data.
The deviation is significant; 
applying higher order polynomial extrapolations to the whole data, 
$\kappa_c$ differs by $10\sigma$,
and the lattice spacing by $6.5\sigma(7\%)$.


\section{Chiral extrapolations based on ChPT}

\begin{figure}[tb]
\vspace{-4mm}
\begin{center}
\scalebox{0.60}{\includegraphics{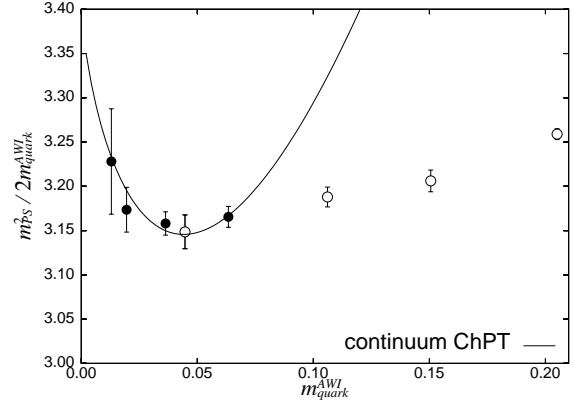}}
\end{center}
\vspace{-13mm}
\caption{
      Simultaneous continuum ChPT fit
      to pseudoscalar meson mass and decay constant.
}
\label{figure:chiral_log_m_quark_AWI_m_PS2_and_renormalized_f_PS}
\vspace{-15pt}
\end{figure}

A choice for controlled chiral extrapolations is to
incorporate chiral perturbation theory (ChPT).
However, the present lattice data are not quite consistent with
ChPT in the continuum~\cite{Spectrum.Nf2},  
which predict the following quark mass dependence to one-loop order:
\begin{eqnarray}
 \frac{m_{PS}^{2}}{2 B_{0} m_{quark}}
 = 1 + \frac{1}{2}
         \frac{2 B_0 m_{quark}}{(4 \pi f)^2}
         \log \frac{2 B_0 m_{quark}}{\Lambda_3^2},\nonumber
\label{equation:chiral_log_m_PS2}
\end{eqnarray}
where $B_0$, $f$ and $\Lambda_3$
are parameters to be obtained by fits.
Fitting the data over the whole range $m_{PS}/m_{V}=0.80$--0.35
yields a large $\chi^2/dof \sim 70$.
We have to restrict the fitting interval to $m_{PS}/m_{V}=0.60$--0.35
to obtain a reasonable fit with $\chi^2 / dof = 1.9$,
as represented
in Fig.~\ref{figure:chiral_log_m_quark_AWI_m_PS2_and_renormalized_f_PS}.
This may be an indication that the chiral logarithm is visible only 
at $m_{PS}/m_{V} \simlt 0.40$.


Another possibility that
we do not clearly observe chiral logarithm in our data
is the suppression by explicit chiral symmetry breaking
of the Wilson quark action; 
modifications of ChPT due to finite lattice spacings
may be needed for analyses of data obtained on coarse lattices.

Recently studies were made to adapt
ChPT to the Wilson-type fermion
at finite lattice spacings (WChPT)~\cite{WChPT.Oliver}. 
Care has to be exercised in the order counting since 
formulae for observables are different.
We employ the WChPT formulae which treat the $O(a^2)$ effects
at the leading order~\cite{WChPT.Sinya},
because the existence
of parity-broken phase and vanishing of pion mass depend on them in a
critical way.

\begin{figure}[tb]
\vspace{-4mm}
\begin{center}
\scalebox{0.60}{\includegraphics{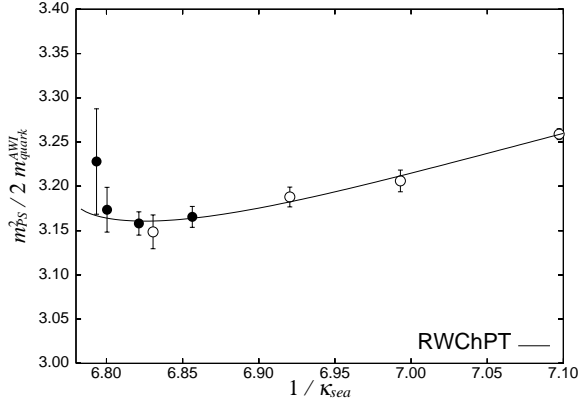}}
\end{center}
\vspace{-13mm}
\caption{
     The resummed WChPT fit
     to pseudoscalar meson mass and AWI quark mass.
}
\label{figure:chiral_log_m_PS2_m_quark_AWI_renormalized_f_PS_WChPT}
\vspace{-15pt}
\end{figure}

There are two features in these formulae worth emphasizing:
(i) the coefficients
of chiral logarithm terms receive $O(a)$ contributions,
modifying the chiral behavior at finite lattice spacings, 
(ii) terms of the form $a^2\log m_{quark}$ appear,
which are more singular
than the $m_{quark}\log m_{quark}$ terms toward the chiral limit
at a finite lattice spacing.
The $a^2\log m_{quark}$ terms must be resummed.
The resummed WChPT formulae (RWChPT) 
proposed by Aoki~\cite{WChPT.Sinya} read,
\begin{eqnarray}
 m_{PS}^2 &=& A m_{quark}^{VWI}
              \left( - \log \left( \frac{A m_{quark}^{VWI}}
                                        {\Lambda_0^2} \right)
              \right)^{\omega_0} \nonumber \\
          &&  \left( 1 + \omega_1^{PS} m_{quark}^{VWI}
                         \log \left( \frac{A m_{quark}^{VWI}}
                                          {\Lambda_3^2} \right)
              \right),
 \label{equation:chiral_log_m_PS2_resummed_WChPT} \nonumber \\
 m_{quark}^{AWI} &=& m_{quark}^{VWI}
                     \left( - \log \left( \frac{A m_{quark}^{VWI}}
                                               {\Lambda_0^2} \right)
                     \right)^{\omega_0} \nonumber \\
            &&       \left( 1 + \omega_1^{AWI} m_{quark}^{VWI}
                                \log \left( \frac{A m_{quark}^{VWI}}
                                                 {\Lambda_{3,AWI}^2} \right)
                     \right),
 \label{equation:chiral_log_m_quark_AWI_resummed_WChPT} \nonumber 
\end{eqnarray}
where the fit parameters are $\kappa_c$ in $m_{quark}^{VWI}$, $A$,
$\omega_0$, $\omega_1^{PS}$, $\omega_1^{AWI}$,
$\Lambda_0$, $\Lambda_3$ and $\Lambda_{3,AWI}$.
As we show in Fig.~\ref{figure:chiral_log_m_PS2_m_quark_AWI_renormalized_f_PS_WChPT},
the one-loop RWChPT formulae is capable of fitting our data
over the whole range $m_{PS}/m_{V}=0.80$--0.35.
The fit result shows that the leading and the one-loop contributions
in RWChPT converges well and
the parameter values are comparable with
phenomenological estimates.
It is also found that if we make a RWChPT extrapolation
using our previous data at $m_{PS}/m_{V}=0.80$--0.55, 
the new data at $m_{PS}/m_{V}=0.60$--0.35 lie well on the extrapolation 
curve.

Encouraged by the features above of the WChPT fit, 
we reanalyzed the previous data
at $m_{PS}/m_{V}=0.80$--0.55 on finer lattices with $a = 0.16(\beta=1.95)$ 
and 0.11~fm ($\beta=2.1$) using WChPT
and performed the continuum extrapolation.
For the quark mass in the continuum limit, we find 
$m_{ud}^{\overline{\rm MS}}(\mu=2~\mbox{GeV}) = 3.11(17)$~[MeV],
which is smaller than the previous result~\cite{CP-PACS.Nf2}
based on the quadratic chiral extrapolation 
by approximately 10\%.


\section{Conclusions}

We have extended the study of the light hadron spectrum
and the quark mass in two-flavor QCD\cite{CP-PACS.Nf2} 
to smaller sea quark masses corresponding to $m_{PS}/m_{V}=0.60$--0.35.
The new data clearly show systematic deviations from the previous
quadratic chiral extrapolation using the data in the range
$m_{PS}/m_{V}-0.80$--0.55.
Whereas fits with either polynomial or continuum
chiral perturbation theory (ChPT) fails, the Wilson ChPT
(WChPT) that includes $a^2$ effects associated with explicit
chiral symmetry breaking successfully fits the whole data.
In particular, WChPT correctly predicts the light quark mass
behavior from data at  medium heavy quark masses such as
$m_{PS}/m_V \simgt 0.5$.

While our initial WChPT analyses gave encouraging results, 
further studies at smaller lattice spacings are needed
to establish that the WChPT parameters show the expected lattice spacing 
dependence.  
 
\vspace{2mm}

Present simulations were performed on VPP-5000 at the Science Information 
Processing Center (SIPC), University of Tsukuba.
This work is supported in part
by Large-Scale Numerical Simulation Project of SIPC,
and by Grants-in-Aid of the Ministry of Education 
(Nos.\ 
12304011, 
12740133, 
13135204, 
13640259, 
13640260, 
14046202, 
14740173, 
15204015, 
15540251, 
16028201
). 



\end{document}